\documentclass[pra,aps,showpacs,twocolumn,floatfix]{revtex4}
\usepackage{graphicx}
\usepackage[ansinew]{inputenc}
\usepackage{array}
\usepackage{color}
\usepackage{amsmath}
\usepackage{amsxtra}
\usepackage{amstext}
\usepackage{amssymb}
\usepackage{latexsym}
\usepackage{dsfont}
\usepackage{verbatim}
\usepackage{comment}

\begin{document}

\title
{Testing the reliability of a velocity definition in dispersive medium}

\author{Mehmet Emre Ta\c{s}g\i n}
\affiliation{Department of Electrical and Electronics Engineering, K{\i}rklareli University, 39060 Kavakl{\i}, K{\i}rklareli, Turkey}

\date{\today}

\begin{abstract}
We introduce a method to test if a given velocity definition  corresponds to an actual physical flow in a dispersive medium. We utilize the equivalence of the pulse dynamics in the real-$\omega$ and real-$k$ Fourier expansion approaches as a test tool. To demonstrate our method, we take the definition introduced by Peatross {\it et al.}  [Phys. Rev. Lett. {\bf 84}, 2370 (2000)] and calculate the velocity in two different ways. We calculate i) the mean arrival time between two positions in space, using the real-$\omega$ Fourier expansion for the fields and ii) the mean spatial displacement between two points in time, using the Fourier expansion in real-$k$ space. We compare the velocities calculated in the two approaches. If the velocity definition truly corresponds to an actual flow, the two velocities must be the same. However, we show that the two velocities differ significantly (3\%) in the region of superluminal propagation even for the successful definition of Peatross {\it et al.}
\end{abstract}


\maketitle

\section{Introduction}
Studies concerning the propagation of light in dispersive media dates back as far as Brillouin and Sommerfeld \cite{brillouin}. Nevertheless, new studies \cite{talukderPRL2005,kohmotoPRE2006,kohmotoPRA2005,talukderPRL2001,talukderPRA2005,oughstunJOA2002,peatrossPRL2000,peatrossJOSAA2001,oughstunJMO2005,nandaPRE2006,
nandaPRA2009,kuzmichPRL2001,zhuyangEPJD2005} on the concept of pulse velocity were stimulated by the famous experiment \cite{wang,chuwong}, where light seems to propagate over the speed of light in vacuum (superluminal). This effect takes place due to superluminal group velocities near the absorption resonance in dye solutions. 

Beside absorptive dielectrics there exist metamaterials where index exhibit some unusual behaviour. These are constructed either using coherent population trapping \cite{EIT}, e.g. Electromagnetically induced transparency (EIT) and index enhancement schemes, or spatial modulation as in left handed materials \cite{Ozbay}. Beyond scientific curiosity, applications such as memory storage/read out using slow light motion carried out the physics of metamaterials to an important status in the information technology.

In such materials, dielectric response may switch from absorption to gain in MHz intervals for the optical frequencies. As well, index of refraction may display steep and negative derivatives. In those situations, it is confusing even to figure out the reorganization of the pulse beside facing the problem of superluminality. Unfortunately, when the original pulse is severely modified there is no direct way to test the validity of the propagation velocity. The experiments \cite{talukderPRL2005,kohmotoPRE2006,kohmotoPRA2005,talukderPRL2001,talukderPRA2005} measure either the peak of the pulse or mean absorption time. However, motion of the pulse peak or center may not correspond to a travel velocity since the shape of the pulse is distorted by mutual act of gain/absorption. For this reason, we adopt a method to test the reliability for a given description of velocity. We utilize the equivalence of the pulse dynamics within the real-$\omega$ and real-$k$ Fourier expansions as the test tool in complex dielectrics.

When the dielectric function is complex, there are two alternative mathematical approaches that are used in dealing with pulse dynamics. One can analyse the system equivalently using the real-$\omega$ $\big(E(x,t)=\int_{-\infty}^{+\infty}d\omega D_1(\omega)e^{i(k(\omega)x-\omega t)}\big)$ and real-$k$ $\big(E(x,t)=\int_{-\infty}^{+\infty}dk D_2(k)e^{i(kx-\omega(k) t)}\big)$ Fourier domain. $E(x,t)$ is the Electric Field that is used to calculate the velocity. In the real-$\omega$ or real-$k$ approaches, decay appears spatially or temporarily in the Fourier integrand, respectively, since $k$ or $\omega$ is complex. If one is interested in the penetration depth, Fourier expansion is carried out over the real-$\omega$ space. On the other hand, if one is interested in the temporal lifetime of the pulse in the material, then real-$k$ expansion is used. In example, dealing with photonic crystals composed of complex dielectric materials, one constructs the Master equation using real-$\omega$ (real-$k$) Fourier space for calculating the penetration (duration) of light into (in) the crystal \cite{maradudinPRB1997,tasginPRA2007,jacksonrealk}. For a given velocity definition, pulse speed can be calculated within both approaches.

Among different velocity definitions \cite{oughstunJOA2002,tanakaPRA1986,smithAJP1970} existing in the literature one of the most successful is the one that is introduced by Peatross {\it et al.} \cite{peatrossPRL2000}. In Ref. \cite{peatrossPRL2000}, propagation is described with Poynting vector average of the temporal position, i.e. $\langle t\rangle_x=\int dt \, t \, S(x,t)/\int dt S(x,t)$. The velocity $v_1=(x-x_0)/(\langle t\rangle_x-\langle t\rangle_{x_0})$ is introduced in this regard. On the other hand, considering the same definition \cite{peatrossPRL2000}, a second velocity $v_2=(\langle x\rangle_t-\langle x\rangle_{t_0})/(t-t_0)$ can be adopted similarly using the mean spatial position of the pulse, i.e. $\langle x\rangle_t=\int dx \, x \, S(x,t)/\int dx S(x,t)$. Since $t$ or $x$ average is dealt within the calculation of $v_1$ or $v_2$, respectively, it is standard to work in the conjugate Fourier space where $\omega$ or $k$ is chosen as real. If the definition correctly addresses a physical flow, then the two velocities must be identical or at least must be very close. 

It is shown \cite{nandaPRE2006} that the observed consistency of the definition \cite{peatrossPRL2000} with the experimental results  \cite{talukderPRL2005,kohmotoPRE2006,kohmotoPRA2005,talukderPRL2001,talukderPRA2005} follows from the equivalence of the detector time (mean time for detector absorption) to the arrival time deduced from this description \cite{peatrossPRL2000}. Accordingly, here we choose to test the validity of this definition as the example.

In order to compare the two results for the given velocity definition, we perform the following. We first calculate the mean arrival time $\Delta t=\langle t\rangle_x$-$\langle t\rangle_{x_0}$, from position $x_0$ to $x$, for a distance $\Delta x_1=x-x_0$. This is handled in the real-$\omega$ approach. Second, we calculate the corresponding mean propagation distance $\Delta x_2=\langle x\rangle_t$-$\langle x\rangle_{t_0}$, from time $t_0$ to $t$, in the real-$k$ Fourier expansion. For the purpose of comparison, we chose $t-t_0$ equal to $\Delta t$ which is the value determined in the real-$\omega$ expansion. Afterwards, we compare the two distances, $\Delta x_1$ and $\Delta x_2$, for the same $\Delta t$. Thereby, we compare the two velocities $v_1=\Delta x_1/\Delta t$ and $v_2=\Delta x_2 /\Delta t$.

Paper is organized as follows. In Sec. II, we establish a connection between the two expansion coefficients $D_1(\omega)$ and $D_2(k)$ using the boundary conditions.  In Sec. III, we calculate the velocity definition of Peatross {\it et al.} \cite{peatrossPRL2000} in two different ways, by expanding the fields both in the real-$\omega$ and real-$k$ Fourier space. We show that there occurs differences in the calculated values of the two velocities, especially in the superluminal region. Sec. IV includes our conclusions.

%%%%%%%%%%%%%%%%%%%%%%%%%%%%%%%%%%%%%%%%%%%%%%%%%%%%%%%%%%%%%%%%%%%%%%%%%%%%%%%%%%%%%%%%%%%%%%%%%%%%%%%%%%%
%%%%%%%%%%%%%%%%%%%%%%%%%%%%%%%%%%%%%%%%%%%%%%%%%%%%%%%%%%%%%%%%%%%%%%%%%%%%%%%%%%%%%%%%%%%%%%%%%%%%%%%%%%%
%%%%%%%%%%%%%%%%%%%%%%%%%%%%%%%%%%%%%%%%%%%%%%%%%%%%%%%%%%%%%%%%%%%%%%%%%%%%%%%%%%%%%%%%%%%%%%%%%%%%%%%%%%%
\section{Relating The Fourier Coefficients $D_1(\omega)$ and $D_2(k)$}

In this section, we establish an analytical connection between the Fourier components of the real-$\omega$ expansion ($D_1(\omega)$) and the real-$k$
expansion ($D_2(k)$). In order to indicate that a variable is fixed to real, we use a bar accent over that variable. For instance, $d\bar{\omega}$ ($d\bar{k}$) corresponds to an integration over the real $\omega$ ($k$) space.

We consider a dielectric function in the Lorentzian form \cite{jackson331,tanakaPRA1986}
\begin{equation}
\epsilon(\omega)=1-\frac{\omega_p^2}{\omega^2-\omega_0^2+i\gamma \omega},
\label{eq:epsilon}
\end{equation}
where $\omega_p$ ($\omega_0$) is the plasma (atomic transition) frequency and $\gamma$ is the damping rate. One can calculate refractive index by $n(\omega)=\left(\epsilon\right)^{1/2}$.
%%
%%
%\begin{figure}
%\includegraphics[width=3.2in]{fig1.eps}
%\caption{Real ($n_R(\omega)$) and imaginary ($n_I(\omega)$) parts of the complex index $n(\omega)=n_R(\omega)+in_I(\omega)$ corresponding to dielectric function in Eq. (\ref{eq:epsilon}) with parameters $\omega_p=0.1\omega_0$ and $\gamma=0.12\omega_0$}.
%\label{fig1}
%\end{figure}
%%
%%

\begin{figure}
\includegraphics[width=3.2in]{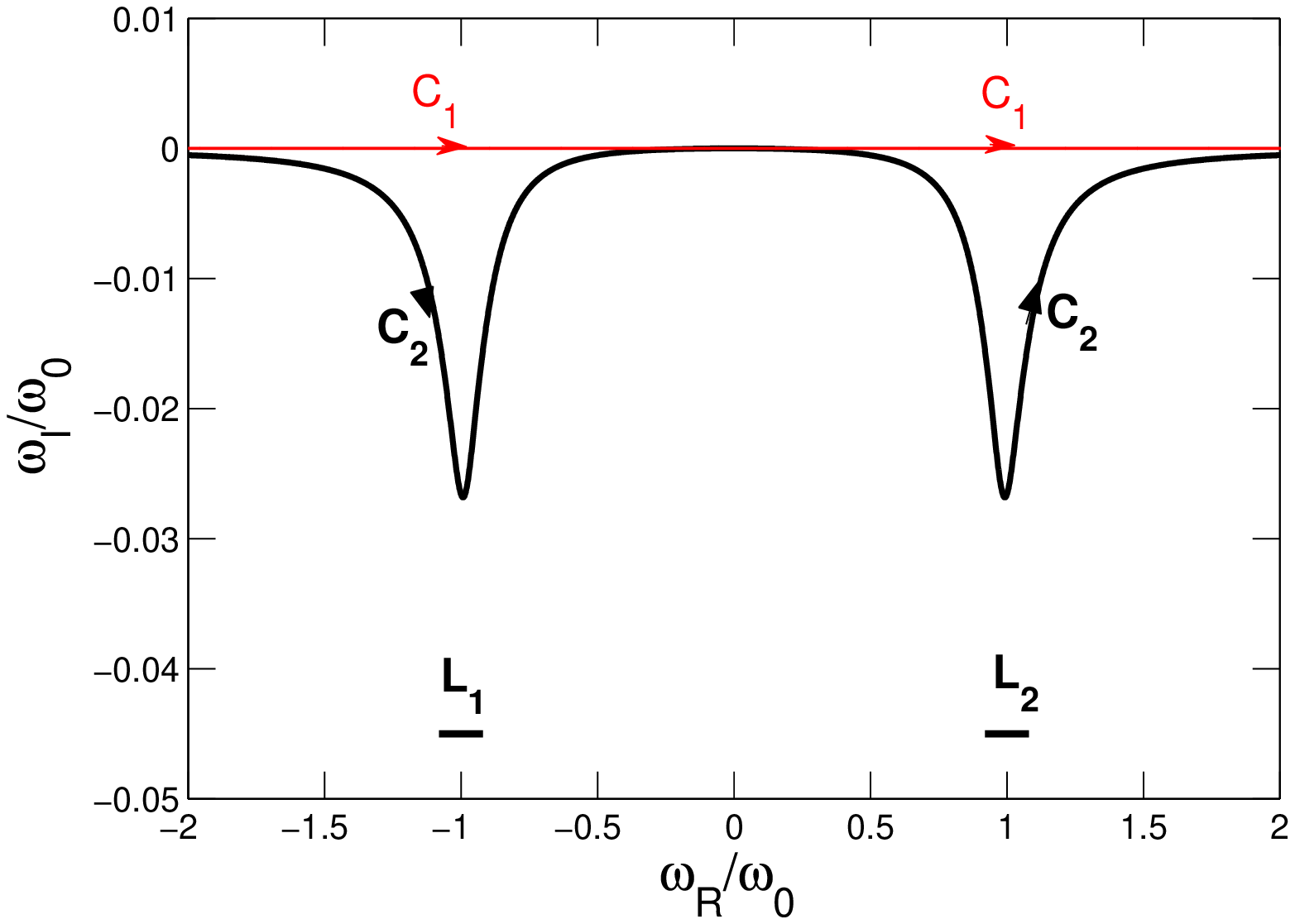}
\caption{(color online) Real-$\omega$ ($C_1$: red thin $\omega_I=0$ line) and real-$k$ ($C_2$: thick contour) integration paths for index $n(\omega)$ plotted for dielectric constant (\ref{eq:epsilon}) with $\omega_p=0.1\omega_0$ and $\gamma=0.12\omega_0$. $C_2$ contour is deduced from the nonlinear index equation $c\bar{k}=n(\omega)\omega$ for the real $\bar{k}$ values. The two lines $L_1$ and $L_2$ are the branch cuts of the index \cite{jackson331}. Both branch cuts are below the real-$k$ integration path. Length of the branch cuts are exaggerated only for the visual purposes.}
\label{fig1}
\end{figure}

The calculation of complex $k$ values for given real $\bar{\omega}$ value is straightforward using $ck=\bar{\omega}n(\bar{\omega})$. However, the calculation of complex $\omega$ requires the solution of the nonlinear equation $c\bar{k}=\omega n(\omega)$ for given $\bar{k}$ real value. In Fig.~\ref{fig1} we plot both integration paths in the complex $\omega$ plane, corresponding to real-$\omega$ ($C_1$ line) and real-$k$ ($C_2$ contour). We also indicate the branch cuts ($L_{1,2}$) of index $n(\omega)$ \cite{jackson331}, which are below the real-$k$ integration path $C_2$.

\begin{figure}
\includegraphics[width=3.8in]{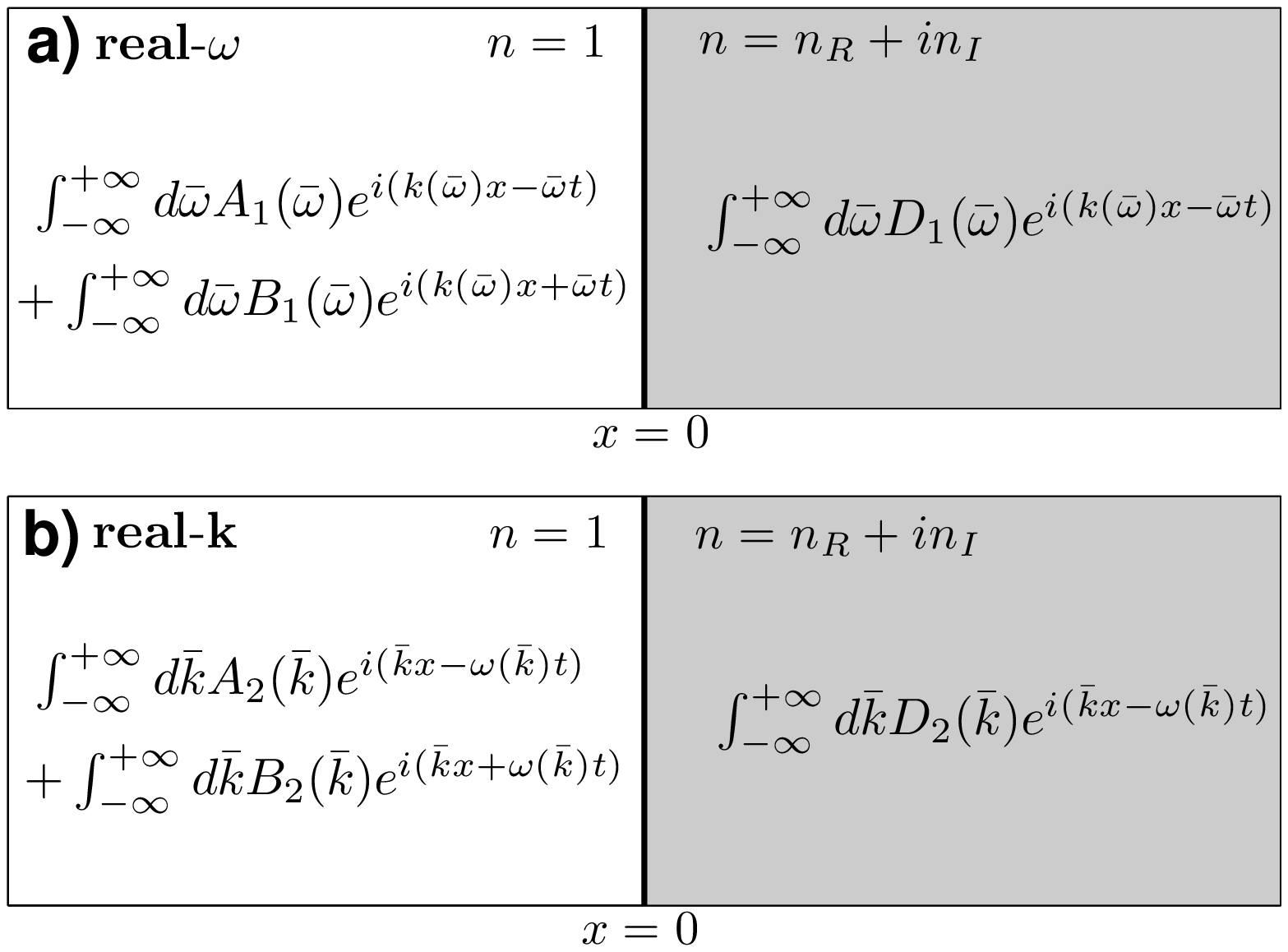}
\caption{The same reflection/transmission problem considered (a) in the real-$\omega$ and (b) in the real-$k$ Fourier spaces. Incident light penetrates from vacuum on the left hand side (LHS) to an absorbing dielectric slab of complex index $n(\omega)$ on right hand side. (a) Incident wave of Fourier coefficients $A_1(\omega)$ results a reflected wave of coefficient $B_1(\omega)$ and a transmitted wave of coefficients $D_1(\omega)$. (b) Incident wave of Fourier coefficient $A_2(k)$ results a reflected wave of coefficient $B_2(k)$ and a transmitted wave of coefficient $D_2(k)$. Since on the LHS index is real and unity, it is simply $A_2(k)=cA_1(\omega)$ and $B_2(k)=cB_2(\omega)$.}
\label{fig2}
\end{figure}

In order to relate Fourier coefficients $D_1(\omega)$ and $D_2(k)$, we consider the following simple reflection/transmission boundary problem (see Fig.~\ref{fig2}). A Gaussian wave packet (of Fourier coefficient $A_1(\omega)$) travelling towards right in vacuum ($n=1$) is incident on the absorbing dielectric slab of index $n(\omega)$, see Fig.~\ref{fig2}a. It results in a reflected wave packet of Fourier coefficient $B_1(\omega)$ and a transmitted  wave (into the slab)
\begin{equation}
E_1(x,t)=\int_{-\infty}^{+\infty} d\omega D_1(\omega) e^{i(k(\omega)x-\omega t)}
\label{eq:D1w}
\end{equation}  
with Fourier coefficient $D_1(\omega)$. In Fig.~\ref{fig2}b, the same problem is considered in the real-$k$ Fourier domain; with incident wave of coefficient $A_2(k)$, reflected wave of coefficient $B_2(k)$, and a transmitted wave
\begin{equation}
E_2(x,t)=\int_{-\infty}^{+\infty} dk D_2(k) e^{i(kx-\omega(k) t)}
\label{eq:D2k}
\end{equation} 
of Fourier coefficient $D_2(k)$.

Since both $\omega$ and $k$ are real on the left hand side, they are simply related by $A_2(k)=cA_1(\omega)$ and $B_2(k)=cB_2(\omega)$. In the real-$\omega$ approach, using the boundary condition (BC) at $x=0$, one obtains
\begin{equation}
D_1(\omega)=\frac{2}{1+n(\omega)} A_1(\omega).
\end{equation}
In this paper, for the sake of simplicity, we consider a Gaussian profile for $A_1(\omega)$ which does not have any pole.

Additionally, the two solutions for the transmitted wave, $E_1(x,t)$ and $E_2(x,t)$, must match at the boundary $x=0$ for all times. That is
\begin{equation}
\int_{-\infty}^{+\infty} d\bar{\omega} D_1(\bar{\omega}) e^{-i\bar{\omega}t} = 
\int_{-\infty}^{+\infty} d\bar{k} D_2(\bar{k}) e^{-i\omega(\bar{k})t} ,
\label{eq:BC}
\end{equation}
where $\bar{\omega}$, $\bar{k}$ stand for real variables and $\omega(\bar{k})=\omega_{R}(\bar{k})+i\omega_{I}(\bar{k})$ is complex function of the real variable $\bar{k}$ determined from the nonlinear index equation $c\bar{k}=\omega n(\omega)$.

Integrating both sides with $\int_{-\infty}^{+\infty} dt e^{i\omega(\bar{k}')}$ in Eq. (\ref{eq:BC}), one obtains
\begin{multline}
 \lefteqn{\int_{\infty}^{+\infty} d\bar{\omega} D_1(\bar{\omega}) \int_{-\infty}^{+\infty} dt e^{i\left(\omega(\bar{k}')-\bar{\omega}\right) t}} %\nonumber
 \\
=\int_{\infty}^{+\infty} d\bar{k} D_2(\bar{k}) \int_{-\infty}^{+\infty} dt  e^{i\left(\omega(\bar{k}')-\omega(\bar{k})\right)t} .
\label{eq:BCdt}
\end{multline} 
Time integrations in Eq. (\ref{eq:BCdt}) are in the form of Dirac-delta function ($\xi(z)$) that is generalized to complex argument \cite{complexdirac}
\begin{equation}
\int_{-\infty}^{+\infty} dt e^{-izt}=2\pi\xi(z),
\label{eq:zdirac}
\end{equation}
with the sampling property \cite{PSzdirac}
\begin{equation}
\mathop{ \int}_{-\infty+i{\rm Im}\{\alpha\}}^{+\infty+i{\rm Im}\{\alpha\}} dz F(z) \xi(z-\alpha) = F(\alpha) ,
\end{equation}
where $\alpha$ and $z$ are complex. Thus, time integrals in Eq. (\ref{eq:BCdt}) result $\xi\left(\omega(\bar{k}')-\bar{\omega})\right)$ and $\xi\left(\omega(\bar{k}')-\omega(\bar{k}))\right)$, respectively. Now, Eq. (\ref{eq:BCdt}) becomes
\begin{multline}
\int_{-\infty}^{+\infty} d\bar{\omega} D_1(\bar{\omega}) \xi\left(\omega(\bar{k}')-\bar{\omega}\right) \\
=\int_{-\infty}^{+\infty} d\bar{k} D_2(\bar{k}) \xi\left(\omega(\bar{k}')-\omega(\bar{k})\right) ,
\label{eq:BCxi}
\end{multline}
where $\bar{\omega}$ is real and $\omega(\bar{k})$ is complex. Note that $d\bar{k}$ corresponds to the $C_2$ contour in Fig.~\ref{fig2} and all possible values of both $\omega(\bar{k}')$ and $\omega(\bar{k})$ rely on the $C_2$ curve. 

Since $D_1(\omega)$ and $\xi\left( \omega(\bar{k}')-\bar{\omega}\right)$ \cite{PSdiracpole} have no pole or branch cut between $C_1$ line and $\omega_I={\rm Im}\{\omega(\bar{k}')\}$ line for any choice of $\bar{k}'$ (Fig.~\ref{fig2}), integration path in the LHS of Eq. (\ref{eq:BCxi}) can be as well written as
\begin{equation}
\mathop{ \int}_{-\infty+{\rm Im}\{\omega(\bar{k}')\}}^{+\infty+{\rm Im}\{\omega(\bar{k}')\}} d\omega D_1(\omega) \xi\left(\omega(\bar{k}')-\omega)\right)= D_1(\omega(\bar{k})).
\label{eq:D1int}
\end{equation}
On the other hand, using $dk=d\omega \frac{dk}{d\omega}$ transformation RHS of Eq. (\ref{eq:BCxi}) can be written as
\begin{multline}
\int_{-\infty}^{+\infty} d\bar{k} D_2(\bar{k}) \xi\left(\omega(\bar{k}')-\omega(\bar{k}))\right) \\
=\int_{C_2} d\omega \frac{dk}{d\omega} D_2(\bar{k}(\omega)) \xi\left(\omega(\bar{k}')-\omega\right),
\label{eq:D2int}
\end{multline}
where $C_2$ is the correspondence of the real-$k$ integration path in the complex $\omega$ plane, depicted in Fig.~\ref{fig2}. $dk/d\omega=n(\omega)+\omega n'(\omega)$ has no pole or branch cut in between the two contours $C_1$ and $C_2$. In addition, we assume that $D_2(\bar{k}(\omega))$ does not have pole or branch cut in between $C_1$ and $C_2$. Afterwards, we confirm this assumption. Therefore, we shift the $C_2$ integration in Eq. (\ref{eq:D2int}) to the $C_1$ contour and obtain
\begin{equation}
\int_{-\infty}^{+\infty} d\omega \frac{dk}{d\omega} D_2(k(\omega)) \xi\left(\omega(\bar{k}')-\omega\right)
=\frac{dk}{d\omega}(\bar{k}')D_2(\bar{k}').
\label{eq:dkdwD2}
\end{equation}
In obtaining the last term of Eq. (\ref{eq:dkdwD2}), in a similar way to Eq. (\ref{eq:D1int}),  we further assume that $D_2(k(\omega))$ has no pole or branch cut between $C_1$ line and $\omega_I={\rm Im}\{\omega(\bar{k}')\}$ line for any choice of $\bar{k}'$.

Finally, Eq. (\ref{eq:BC}) transforms to the simple relation 
\begin{equation}
D_2(\bar{k})=\frac{d\omega}{dk}(\bar{k}) D_1(\omega(\bar{k})) ,
\label{eq:D1D2}
\end{equation}
where presence of $\bar{k}$ indicates that all quantities are evaluated at complex variable $\omega(\bar{k})$ corresponding to real $\bar{k}$. Eq. (\ref{eq:D1D2}) confirms that $D_2(k)$ indeed does not have pole or branch cut in between the contours $C_1$, $C_2$ and $\omega_I={\rm Im}\{\omega(\bar{k})\}$ line.

Relation (\ref{eq:D1D2}) is quite simple and straightforward. On the other hand, however, result is cumbersome when $D_1(\omega)$ or $n(\omega)$ has a branchcut/pole in between the curves $C_1$ and $C_2$ (see Fig.~\ref{fig1}).
 
In the following section, we use Eq. (\ref{eq:D1D2}) and relate the real-$k$ integrand to the real-$\omega$ one.

%%%%%%%%%%%%%%%%%%%%%%%%%%%%%%%%%%%%%%%%%%%%%%%%%%%%%%%%%%%%%%%%%%%%%%%%%%%%%%%%%%%%%%%%%%%%%%%%%%%%%%%%%%%
%%%%%%%%%%%%%%%%%%%%%%%%%%%%%%%%%%%%%%%%%%%%%%%%%%%%%%%%%%%%%%%%%%%%%%%%%%%%%%%%%%%%%%%%%%%%%%%%%%%%%%%%%%%
%%%%%%%%%%%%%%%%%%%%%%%%%%%%%%%%%%%%%%%%%%%%%%%%%%%%%%%%%%%%%%%%%%%%%%%%%%%%%%%%%%%%%%%%%%%%%%%%%%%%%%%%%%%
\section{Comparison of Velocities in the Real-$\omega$ and the Real-$k$ Approaches}

\begin{figure}
\includegraphics[width=3.2in]{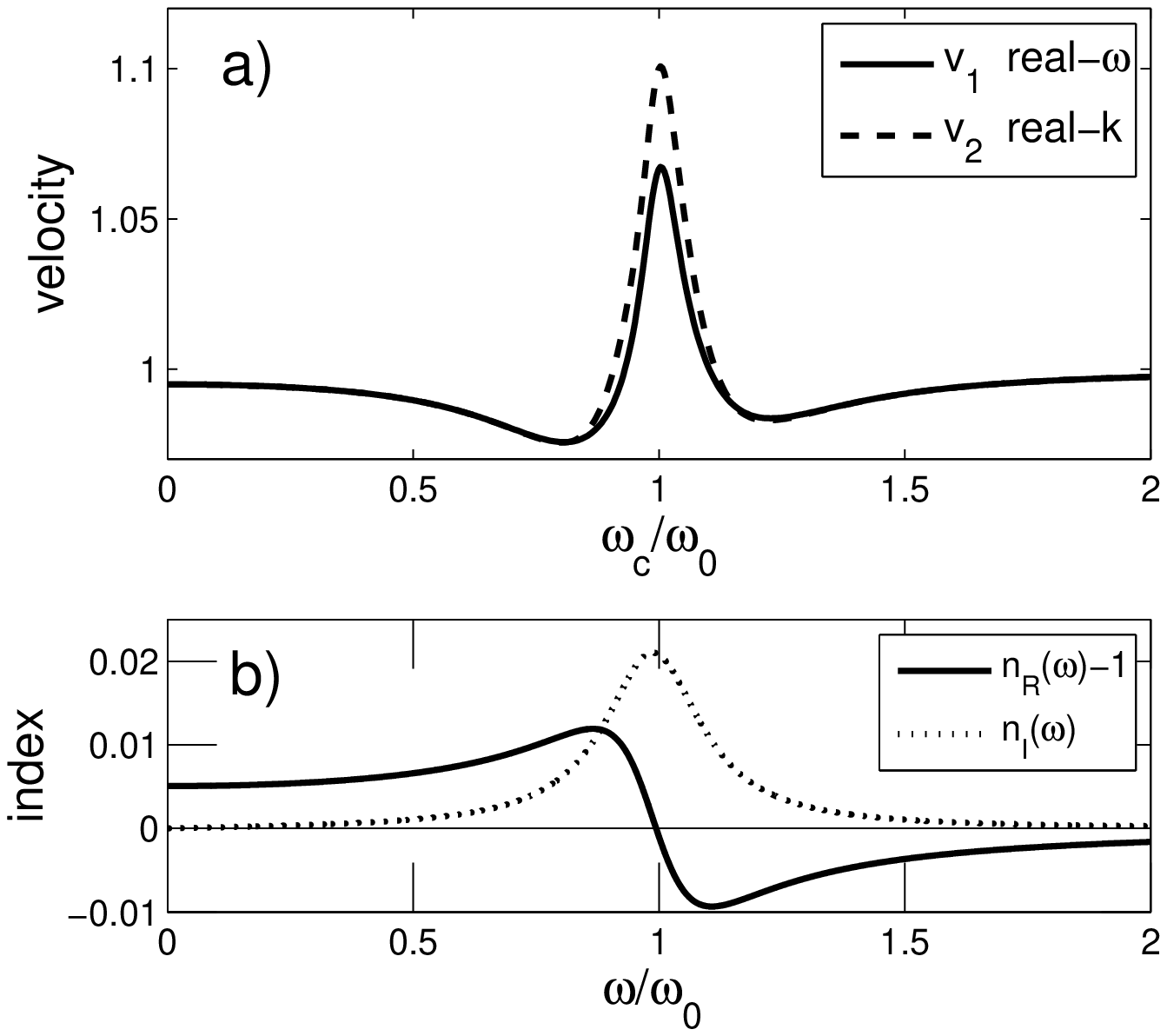}
\caption{(a) Comparison of the two velocities (in units of c) deduced from definition of  Peatross. {\it et al.} \cite{peatrossPRL2000}. The velocities $v_1=\Delta x/\big(\langle t\rangle_{\Delta x}-\langle t\rangle_0\big)$  and $v_2=\big(\langle x\rangle_{\Delta t}-\langle x\rangle_0\big)/\Delta t$  are calculated using the same definition, but performing real-$\omega$ (solid line) and real-$k$ (dashed line) Fourier expansions for the fields, respectively. 
For a consistent definition, the two results must be identical. However, 3\% discrepancy between $v_1$ and $v_2$ is observed in the superluminal region. Thus, this description is not so reliable in the superluminal regime. We use a Gaussian pulse of carrier frequency $\omega_c$ and temporal width $\omega_0\tau=20$. Propagation distance is $\Delta x_1=150\, c/\omega_0$. (b) Real ($n_R$) and imaginary ($n_I$) parts of the index of refraction. Parameters for the index are the same with Fig.~\ref{fig1}. }
\label{fig3}
\end{figure}

In this section, we test the velocity definition introduced by Peatross {\it et al.} \cite{peatrossPRL2000}. We derive the two expressions for the velocity, using the real-$\omega$ and real-$k$ Fourier expansions. Afterwards, we calculate the velocities $v_1$ and $v_2$ by relating the coefficients $D_1(\omega)$ and $D_2(k)$ (see Eq.s (\ref{eq:D1w},\ref{eq:D2k})) using Eq. (\ref{eq:D1D2}). In the superluminal region, we observe discrepancies in the amount of 3\% (see Fig.~\ref{fig3}a).

In Sec. II we derive the relation (\ref{eq:D1D2}) imagining a boundary between vacuum and dielectric along the lines of experiments \cite{talukderPRL2005,kohmotoPRE2006,kohmotoPRA2005,talukderPRL2001,talukderPRA2005} and the generality. In this section, however, we consider a more particular case; a dispersive dielectric occupying the whole space without any boundaries. This is done so, not to deal with the tail of the pulse relying out of the dielectric. Relations (\ref{eq:BC}) and (\ref{eq:D1D2}) remain valid. Because, the condition of matching the two solutions at the origin $E_1(0,t)=E_2(0,t)$ does not require a physical boundary. At any random position this condition must already be satisfied.

We consider a Gaussian wave packet $U(0,t)=e^{-t^2/\tau^2}\text{cos}(\omega_ct)$ imposed at the origin. This leads \cite{tanakaPRA1986}  to the Fourier coefficient $D_1(\omega)=e^{-(\omega-\omega_c)^2/4}+e^{-(\omega+\omega_c)^2/4}$, where $\omega_c$ is the carrier frequency of the pulse. $D_2(k)$ is determined from $D_1(\omega)$ using Eq. (\ref{eq:D1D2}).

First, we calculate the arrival time ($\Delta t=\langle t\rangle_{\Delta x_1}-\langle t\rangle_{0}$) of the mean pulse center from position $0$ to $\Delta x_1$. Time averages are directly calculated within the real-$\omega$ expansion of the fields similar to Ref. \cite{peatrossPRL2000}, using Fourier coefficient $D_1(\omega)$. Second, we use the same arrival time $\Delta t$ (which is calculated using real-$\omega$ domain) in the real-$k$ approach and evaluate the displacement of the average pulse position, i.e. $\Delta x_2=\langle x\rangle_{\Delta t}-\langle x\rangle_0$, from time $0$ to $\Delta t$. Finally, since $\Delta t$ is common in both approaches, we compare the velocities $v_1=\Delta x_1/\Delta t$ and $v_2=\Delta x_2/ \Delta t$ plotted in Fig.~\ref{fig3}a. 

%%%%%%%%%%%%%%%%%%%%%%%%%%%%%%%%%%%%%%%%%%%%%%%%%%%%%%%%%%%%%%%%%%%%%%%%%%%%%%%%%%%%%%%%%%%%%%%%%%%%%%%%%%%
\subsection{Real-$\omega$}
Average time-position of the pulse after propagating a distance $\Delta x$ (starting from $x=0$)
\begin{equation}
\langle t \rangle_{\Delta x}= \frac{\int dt \, t \, S(\Delta x,t)}{\int dt S(\Delta x,t)},
\label{eq:tSav}
\end{equation}
can be directly calculated by Fourier expansion in real-$\omega$ space using \cite{peatrossPRL2000}
\begin{eqnarray}
\int dt \, t \, S(\Delta x,t) = \Delta x \int_{-\infty}^{+\infty} d\bar{\omega} \frac{dk}{d\omega} e^{-2k_I\Delta x} \left|D_1(\bar{\omega})\right|^2 n^*(\bar{\omega}) \nonumber \\
-i\int_{-\infty}^{+\infty} d\bar{\omega} e^{-2k_I\Delta x} \frac{dD_1}{d\omega} D_1^*(\bar{\omega}) n^*(\bar{\omega}) \qquad
\end{eqnarray}
and
\begin{equation}
\int dt S(\Delta x,t)= \int_{-\infty}^{+\infty} d\bar{\omega} e^{-2k_I\Delta x} \left|D_1(\bar{\omega})\right|^2 n^*(\bar{\omega}).
\end{equation}
Here $k_I(\bar{\omega})$ is the imaginary part of the wave-vector corresponding to the real $\bar{\omega}$ value. The calculated values of the velocity $v_1=\Delta x_1/\big(\langle t\rangle_{\Delta x_1}-\langle t\rangle_0\big)$ for different carrier frequencies $\omega_c$ are plotted in Fig.~\ref{fig3}a with solid line. We choose a temporal width of $\omega_0\tau=20$ and propagation distance of $\Delta x_1=150\, c/\omega_0$.
%%%%%%%%%%%%%%%%%%%%%%%%%%%%%%%%%%%%%%%%%%%%%%%%%%%%%%%%%%%%%%%%%%%%%%%%%%%%%%%%%%%%%%%%%%%%%%%%%%%%%%%%%%%

\subsection{Real-$k$}
Above, using the real-$\omega$ approach, we determine the arrival time $\Delta t$ in between the two positions $0$ and $\Delta x_1$. In the real-$k$ approach, we use the calculated value of $\Delta t$ as the input. We determine the distance that  mean pulse center travels from time $0$ to $\Delta t$, i.e. $\Delta x_2=\langle x\rangle_{\Delta t}-\langle x\rangle_0$. The average pulse position at time $\Delta t$
\begin{equation}
\langle x \rangle_{\Delta t}= \frac{\int dx \, x \, S(x,\Delta t)}{\int dx S(x,\Delta t)},
\label{eq:xSav}
\end{equation}
can be directly calculated by carrying the Fourier expansion over the real-$k$ coefficients using the expressions
\begin{eqnarray}
\int dx \, x \, S(x,\Delta t) &= \Delta t \int_{-\infty}^{+\infty} d\bar{k} \frac{d\omega}{dk} e^{2\omega_I\Delta t} \left|D_2(\bar{k})\right|^2 n^*(\bar{k}) \nonumber \\
& +i\int_{-\infty}^{+\infty} d\bar{k} e^{2\omega_I\Delta t} \frac{dD_2}{dk} D_2^*(\bar{k}) n^*(\bar{k})
\end{eqnarray}
and
\begin{equation}
\int dx S(x,\Delta t)= \int_{-\infty}^{+\infty} d\bar{k} e^{2\omega_I\Delta t} \left|D_2(\bar{k})\right|^2 n^*(\bar{k}) ,
\end{equation}
where Fourier components decay in time with imaginary part ($\omega_I$) of the complex frequency $\omega(\bar{k})$ during the propagation. Complex $\omega(\bar{k})$ values are determined from the nonlinear index equation $c\bar{k}=\omega n(\omega)$ for real $\bar{k}$, and $\omega_I$ is always negative for the absorbing dielectric (\ref{eq:epsilon}). Relevance of Fourier coefficient $D_2(k)$ to $D_1(\omega)$ is given in Eq. (\ref{eq:D1D2}).

The average displacement $\Delta x_2$ that is calculated in the real-$k$ approach is compared with the one for the real-$\omega$ approach $\Delta x_1$. We note that $\Delta t$ is common to both approaches. The calculated velocity $v_2=\big(\langle x\rangle_{\Delta t}-\langle x\rangle_0\big)/\Delta t$ is plotted in Fig.~\ref{fig3}a with dotted line for different carrier frequencies. The two results, $v_1$ and $v_2$, differ significantly (3\%) in the superluminal propagation regime.

On the other hand, similar calculations using the real part of the conventional group velocity \cite{jacksonrealk}, as $v_1={\rm Re}\left\{ d\omega/dk \right\}$ and $v_2=1/{\rm Re}\left\{ dk/d\omega \right\}$, results in 16\% discrepancy in the superluminal propagation region.
%\begin{figure}
%\includegraphics[width=3.2in]{fig5.eps}
%\caption{Same with Fig.~\ref{fig3} but real part of the conventional definition of group velocity is used in the calculations. This definition is less successful compared to Peatross {\it et al.}, since 16\% discrepancy occurs in the superluminal region.}
%\label{fig5}
%\end{figure}

%%%%%%%%%%%%%%%%%%%%%%%%%%%%%%%%%%%%%%%%%%%%%%%%%%%%%%%%%%%%%%%%%%%%%%%%%%%%%%%%%%%%%%%%%%%%%%%%%%%%%%%%%%%
%%%%%%%%%%%%%%%%%%%%%%%%%%%%%%%%%%%%%%%%%%%%%%%%%%%%%%%%%%%%%%%%%%%%%%%%%%%%%%%%%%%%%%%%%%%%%%%%%%%%%%%%%%%
%%%%%%%%%%%%%%%%%%%%%%%%%%%%%%%%%%%%%%%%%%%%%%%%%%%%%%%%%%%%%%%%%%%%%%%%%%%%%%%%%%%%%%%%%%%%%%%%%%%%%%%%%%%

\section{Summary and Conclusions}

The velocity introduced by keeping track of the pulse peak or the pulse center does not always corresponds to the velocity of the energy/signal transfer. When the pulse shape is modified during the propagation, it is confusing even conceptually to define the arrival time of the original signal.

Here, we introduce a method to check if a given velocity definition is reliable regarding its correspondence to a real physical flow. We calculate the velocity introduced by Peatross {\it et al.} \cite{peatrossPRL2000} in two different ways. First, we calculate the mean arrival time $\Delta t$ of the pulse between two points in space. We perform this calculation using real-$\omega$ Fourier expansion of the fields. Second, we calculate the mean displacement of the pulse between two points in time, $0$ and $\Delta t$. This calculation is carried out with real-$k$ Fourier expansion. Finally, since $\Delta t$ is common in the both approaches, we compare the two velocities. 

We observe that the velocity definition of Peatross {\it et al.}, relying on the Poynting vector average of the pulse, results in 3\% discrepancy in the superluminal propagation region, see Fig.~\ref{fig3}a. Thus, one questions if this velocity truly corresponds to a physical flow in the superluminal region. On the other hand, definition of Peatross {\it et al.} is more successful compared to the conventional definition of group velocity, where discrepancy comes out to be 16\% in the superluminal region.

Since the arrival time introduced by Peatross {\it et al.} also corresponds to the detector time \cite{nandaPRE2006}, we reach the additional conclusion that the arrival time measurements \cite{talukderPRL2005,kohmotoPRE2006,kohmotoPRA2005,talukderPRL2001,talukderPRA2005} do not address a proper velocity for the flow. It is still an open problem to find a reliable velocity description, consistent with equivalence of the two approaches. On the other hand, our method is also possible to address the physics of elementary particles when there exist sources standing for absorption/gain \cite{neutrinos}.

%\begin{acknowledgements}
%I gratefully thank G\"{u}rsoy Akk\"{u}\c{c} for his intensive help with the manuscript. I specially thank to Hasan Cortan and Do\v{g}an \c{C}elik for their motivational support.
%\end{acknowledgements}

%%%%%%%%%%%%%%%%%%%%%%%%%%%%%%%%%%%%%%%%%%%%%%%%%%%%%%%%%%%%%%%%%%

%%%%%%%%%%%%%%%%%%%%%%%%%%%%%%%%%%%%%%%%%%%%%%%%%%%%%%%%%%%%%%%%%%
%%%%%%%%%%%%%%%%%%%%%%%%%%%%%%%%%%%%%%%%%%%%%%%%%%%%%%%%%%%%%%%%%%
%%%%%%%%%%%%%%%%%%%%%%%%%%%%%%%%%%%%%%%%%%%%%%%%%%%%%%%%%%%%%%%%%%%
%\begin{appendix}
%\section{Equations of Motion}
%
%
%
%\end{appendix}
%%%%%%%%%%%%%%%%%%%%%%%%%%%%%%%%%%%%%%%%%%%%%%%%%%%%%%%%%%%%%%%%%%
%%%%%%%%%%%%%%%%%%%%%%%%%%%%%%%%%%%%%%%%%%%%%%%%%%%%%%%%%%%%%%%%%%

\begin{thebibliography}{99}


\bibitem{brillouin} L. Brillouin, {\it Wave Propagation and Group Velocity} (Academic Press, New York, 1960).


%Experimental
\bibitem{talukderPRL2005} A.I. Talukder, T. Haruta, and M. Tomita, Phys. Rev. Lett. {\bf 94}, 223901 (2005).
%Propagation of femtosecond light pulses in a dye solution: Nonadherence to the conventional group velocity
\bibitem{kohmotoPRE2006} T. Kohmoto, Y. Fukui, S. Furue, K. Nakayama, and Y. Fukuda, Phys. Rev. E {\bf 74}, 056603 (2006).
\bibitem{kohmotoPRA2005} T. Kohmoto, H. Tanaka, S. Furue, K. Nakayama, M. Kunitomo, and Y. Fukuda, Phys. Rev. A {\bf 72}, 025802 (2005).
\bibitem{talukderPRL2001} Md.A.I. Talukder, Y. Amagishi, and M. Tomita, Phys. Rev. Lett. {\bf 86} 3546 (2001).
\bibitem{talukderPRA2005} A.I. Talukder and M. Tomita, Phys. Rev. A {\bf 72}, 051802(R) (2005).


%Dispersive pulse dynamics and associated pulse velocity measures
\bibitem{oughstunJOA2002} K.E. Oughstun and N.A. Cartwright, J. Opt. A: Pure Appl. Opt. {\bf 4}, S125 (2002).
%Average Energy Flow of Optical Pulses in Dispersive Media
\bibitem{peatrossPRL2000} J. Peatross, S.A. Glasgow, and M. Ware, Phys. Rev. Lett. {\bf 84}, 2370 (2000).

%The Role of the Instantaneous Spectrum on Pulse Propagation in Causal Linear Dielectrics
\bibitem{peatrossJOSAA2001} J. Peatross, M. Ware, and S.A. Glasgow, , J. Opt. Soc. of Am. A {\bf 18}, 1719-1725 (2001). 
%Physical significance of the group velocity in dispersive, ultrashort gaussian pulse dynamics
\bibitem{oughstunJMO2005} K.E. Oughstun and N.A. Cartwright, J. Mod. Opt. {\bf 52}, 1089 (2005).
%Delay times and detector times for optical pulses traversing plasmas and negative refractive media
\bibitem{nandaPRE2006} Lipsa Nanda, Aakash Basu, and S.A. Ramakrishna, Phys. Rev. E 74, 036601 (2006).
%Why do superluminal pulses become subluminal once they go far enough?
\bibitem{nandaPRA2009} L. Nanda, H. Wanare, and S.A. Ramakrishna, Phys. Rev. A {\bf 79}, 041806(R) (2009).
%SNR 
\bibitem{kuzmichPRL2001} A. Kuzmich, A. Dogariu, L.J. Wang, P.W. Milonni, and R.Y. Chiao,  Phys. Rev. Lett. {\bf 86} 3925 (2001).
\bibitem{zhuyangEPJD2005} S.-Y. Zhu, L.-G.Wang, N.-H. Liu, and Y.-P. Yang, Eur. Phys. J. D {\bf 36}, 129 (2005).


\bibitem{wang} L.J. Wang, A. Kuzmich, and A. Dogariu, Nature (London) {\bf 406}, 277 (2000).
\bibitem{chuwong} S. Chu and S. Wong, Phys. Rev. Lett. {\bf 48}, 738 (1982).





%EIT like ....
\bibitem{EIT} M.O. Scully and M.S. Zubairy, {\it Quantum Optics} (Cambridge
University Press, Cambridge, U.K., 1997).

\bibitem{Ozbay} E. Cubukcu, K. Aydin, E. Ozbay, S. Foteinopoulou and C.M. Soukoulis, Nature {\bf 423}, 604 (2003). 

\bibitem{maradudinPRB1997} V. Kuzmiak and A.A. Maradudin, Phys. Rev. B {\bf 55}, 7427 (1997).
\bibitem{tasginPRA2007} M.E. Ta\c{s}g{\i}n, \"{O}.E. M\"{u}stecapl{\i}o\v{g}lu and M.\"{O}. Oktel, Pys. Rev. A {\bf 75}, 063627 (2007).
\bibitem{jacksonrealk} J.D. Jackson, {\it Classical Electrodynamics} (Wiley, New York, $\text{3}^{\text{rd}}$ Ed.  1998) Sec.s 7.8 and 7.9. 

%O.A. Kocharovskaya and Ya.I. Khanin, Pis'ma Zh.\'{E}ksp. Teor.
%Fiz. {\bf 48}, 581 (1988) [JETP Lett. 48, 630 (1988)]; S.E. Harris,
%Phys. Rev. Lett. {\bf 62}, 1033 (1989); M.O. Scully, S.-Y. Zhu, and
%A. Gavrielides, {\it ibid}. {\bf 62}, 2813 (1989); A.S. Zibrov {\it et al}., {\it ibid}.
%{\bf 75}, 1499 (1995); G.G. Padmabandu {\it et al}., {\it ibid}. {\bf 76}, 2053
%(1996).


\bibitem{tanakaPRA1986} M. Tanaka, M. Fujiwara, and H. Ikegami, Phys. Rev. A {\bf 34}, 4851 (1986).
%The velocities of Light
\bibitem{smithAJP1970} R.L. Smith, Am. J. Phys. {\bf 38}, 978 (1970).

%
\bibitem{jackson331} J.D. Jackson, {\it Classical Electrodynamics} (Wiley, New York, $\text{3}^{\text{rd}}$ Ed.  1998)  p.s 331,337.

\bibitem{complexdirac} M.J. Corinthios, IEE Proc.-Vis. Image Signal Process. {\bf 150}, 69 (2003); M.J. Corinthios, IEE Proc.-Vis. Image Signal Process. {\bf 152}, 97 (2005).
\bibitem{PSzdirac} We perform the $\alpha \to iz$ transformation in the Ref. \cite{complexdirac}.
\bibitem{PSdiracpole} Definition of $\xi(z)=\lim_{\tau\to\infty}\int_{-\tau}^{+\tau}e^{-izt}=\lim_{\tau\to\infty}2\sin(z\tau)/z$ results no pole.


%\bibitem{tasginArxiv} M.E. Ta\c{s}g{\i}n, arXiv:1204.5460v2 (2012).

\bibitem{neutrinos} 
http://www.nature.com/news/neutrino-experiment-replicates-faster-than-light-finding-1.9393

\end{thebibliography}
\end{document}